# Optical Studies of Zero-Field Magnetization of CdMnTe Quantum Dots: Influence of Average Size and Composition of Quantum Dots

T. Gurung, S. Mackowski*, H.E. Jackson, and L.M. Smith

*Department of Physics, University of Cincinnati, 45221-0011 Cincinnati OH, USA*

W. Heiss

*Institut für Festkörperphysik, Universität Linz, 4040 Linz, Austria*

J. Kossut and G. Karczewski

*Institute of Physics Polish Academy of Sciences, Al. Lotnikow 32/46, 02-668 Warsaw, Poland*

We show that through the resonant optical excitation of spin-polarized excitons into CdMnTe magnetic quantum dots, we can induce a macroscopic magnetization of the Mn impurities. We observe very broad (4 meV linewidth) emission lines of single dots, which are consistent with the formation of strongly confined exciton magnetic polarons. Therefore we attribute the optically induced magnetization of the magnetic dots results to the formation of spin-polarized exciton magnetic polarons. We find that the photo-induced magnetization of magnetic polarons is weaker for larger dots which emit at lower energies within the QD distribution. We also show that the photo-induced magnetization is stronger for quantum dots with lower Mn concentration, which we ascribe to weaker Mn-Mn interaction between the nearest neighbors within the dots. Due to particular stability of the exciton magnetic polarons in QDs, where the localization of the electrons and holes is comparable to the magnetic exchange interaction, this optically induced spin alignment persists to temperatures as high as 160 K.

* corresponding author, electronic mail: seb@physics.uc.edu

Utilizing the spin degree of freedom in electronic devices offers potential advantages in comparison with conventional charge-based semiconductor applications, such as non-volatility, increase of data processing speed and integration density, and decrease of power consumption [1]. However, in order to build such spin-based devices, one has to be able to create, control and manipulate the spin polarization of carriers in semiconductors. In this regard, diluted magnetic semiconductors (DMS) are considered a promising material system for using the electron spin in future technology [2,3]. Recently, carrier-induced ferromagnetism has been observed in GaMnAs epitaxial layers (where Mn acts as an acceptor), with Curie temperature as high as 150K [4]. On the other hand, in the case of p-doped CdMnTe quantum wells (QWs) such a global ferromagnetic order of manganese impurities has also been obtained, but only at temperatures below 5K [5]. In both these semiconductor systems the ferromagnetic alignment of Mn ions is stabilized through the spin exchange interaction between heavy holes in the valence band and the localized spins of Mn impurities. In fact, this exchange interaction, which is characteristic for DMS materials, may be of the order of 1 eV [6], thus dominating the exciton binding energy, thermal energy, and even quantum confinement in semiconductor heterostructures.

One of the most remarkable effects, which originate from this large spin interaction is the formation of exciton magnetic polarons (EMPs). An EMP is formed when a photo-created exciton spontaneously polarizes magnetic impurities within its Bohr radius, which significantly lowers its energy [7]. It is important to note, that although the Mn ions within each single EMP are aligned ferromagnetically, the direction of magnetization is usually randomly distributed across the sample. In this case no net global magnetization is observed. One may however expect that it might be possible to induce non-zero magnetization in a DMS material by photo-excitation of spin-polarized excitons.

Such an optically induced magnetization has been observed previously in bulk DMS samples [8,9], DMS QWs [10], and most recently in magnetically doped self-assembled QDs [11]. In each case, this spin-polarization of EMPs has only been seen for optical excitation of excitons into localized electronic states. For bulk materials or QW structures such exciton states exist only in the disordered regions below the mobility edge, and are weakly bound. As a result, the magnetization of EMPs has only been measured using a micro-SQUID magnetometry. Moreover, due to the relatively weak localization of excitons on the width fluctuations of the QW (i.e. instability of EMPs) and the very rapid time decay of the spin alignment [10], this optically induced magnetization exists only at relatively low temperatures (< 6K). It is also important to note that even though a small magnetization was detected by a micro-SQUID magnetometer, no corresponding optical signature for the magnetization was observed, as the decay of the spin alignment has been found to be much faster than the exciton recombination time [12].

Only relatively recently using epitaxial growth it has become possible to incorporate magnetic ions into semiconductor QDs [13-16]. In contrast to bulk materials and QWs, for QDs, the excitons are already strongly confined in all three dimensions to a size comparable to their Bohr radius. Therefore, as it has been recently shown, the optically induced magnetization is strongly enhanced in these QD structures and persists to temperature higher than 120 K [11]. We believe that the strongly enhanced magnetization of the EMPs in DMS QDs is due to a larger EMP binding energy resulting from the strong electronic confinement of excitons [11]. An additional advantage of such strong confinement is a suppression of the spin scattering processes of photo-excited excitons [17]. Moreover, it has also been observed in CdMnSe QDs, that the three-dimensional confinement reduces the formation time of the EMP in magnetic QDs [18].

It is well known from the extensive work on bulk DMS [19-21] and DMS-based QW structures [22-23] that the magnetic properties of the EMPs depend on the concentration of magnetic impurities [19] and on the characteristic length of confining potentials [21]. In this work, we extend the measurements reported in Ref. 11 on self-assembled CdMnTe QDs and focus particularly on the effects of Mn-concentration and QD size on the photo-induced magnetization. The ferromagnetic alignment of Mn impurities confined in QDs is probed by resonantly excited polarized photoluminescence (PL) spectroscopy. We study three samples with varying the amount of Mn ions incorporated into the QDs during the growth of the structures. Moreover, we apply rapid thermal annealing in order to change the average QD size and Mn concentration in the studied samples [24,25]. The QD size selectivity is achieved by performing the experiment for different excitation energies within the broadened PL emission [25], as the emission energy is, in the first approximation, related to the QD size. For all three QD samples we find a non-zero polarization of the QD emission at B=0T, which is identical to the polarization of the excitation, which demonstrates that the resonant excitation creates a macroscopic spin polarization of Mn impurities in CdMnTe QDs [11].

The results show that for each QD sample, the polarization of the QD emission is larger for QDs with higher ground state energies, i.e. presumably with smaller average sizes. As this photo-induced magnetization is a result of the formation of spin-aligned EMPs, this may suggest that the polarons are more stable in smaller QDs [26]. We also observe almost two times stronger polarization of the emission for the ensemble of QDs with smaller average Mn concentration. We ascribe this effect to the reduced contribution of the Mn-Mn antiferromagnetic interaction for QDs with smaller Mn content. We believe that presented results provide important insights into the mechanism of the robust optically induced spin alignment of Mn ions in QDs. Furthermore, these experimental results, in addition to already existing theoretical studies [27], should enable to develop a comprehensive theoretical description of this effect.

The samples containing magnetic CdMnTe QDs were grown on a GaAs substrate (001) by molecular beam epitaxy at 350°C. In order to incorporate Mn ions into the sample, the top surface of the thick ZnTe buffer layer was exposed to a Mn flux for several seconds before depositing the 4 monolayer thick CdTe QD layer. Finally, a 50nm-thick ZnTe cap layer was grown over the dot layer. The magnetic CdMnTe QDs were formed by self-assembly. As determined previously by PL measurements in magnetic field [25], the excitons confined to these QDs strongly interact with magnetic ions confined within their Bohr radius: strong polarization of the PL in applied magnetic field has been observed. Two of the samples were grown with different Mn exposure time of 2 and 4.5 seconds with approximate average Mn concentration of 2% and 5%, respectively. In addition, the latter sample was annealed in an Ar environment for 15s at 420°C [25].

Photoluminescence (PL) measurements using both resonant (dye laser, Rhodamine 6G) and non-resonant (514.5nm line of Ar+ laser) excitations were performed to examine the optical properties of CdMnTe QDs. The experiments with resonant excitation were carried out for several laser energies within the inhomogenously broadened QD emission line [25] in order to probe the polarization properties of QDs with different sizes. For macroscopic optical measurements, where a large number of QDs can be probed, an 80-mm focal length lens was used. On the other hand, in order to study the optical properties of single magnetic QDs, a 50X (0.5 NA) microscope objective focused the laser down to ~ 1.7 μm diameter spot. For the latter experiments, magnetic fields up to 4T were applied in the Faraday configuration. Equivalent sets of Babinet-Soleil compensators and Glan-Thompson linear polarizers were used to control the polarization of the excitation, as well as to analyze the polarization of the PL emission. The sample was placed on the copper cold finger of a temperature-controlled continuous-flow helium cryostat. The measurements were carried out for temperatures ranging from 6K to 120K. The signal was detected by a CCD detector after being dispersed by a DILOR triple monochromator with a resolution of 80 µeV.

Incorporation of magnetic ions in semiconductor QDs is known to strongly modify the optical properties of these structures. Apart from enhancing the effects of external magnetic field observed for large QD ensemble [14,15,25] and single QDs [25,28,29], the spin interaction between localized Mn ions and localized electron-hole pairs influences also the zero-field behavior of magnetic QDs. One of the most striking is the broadening of a single magnetic dot emission line [28,30,31].

In Fig. 1 we show a PL emission line at T=6K of large ensemble of CdMnTe QDs ensemble (approximately $10^5$ dots in a 30-micron diameter area) excited above the ZnTe barrier. A complicated spectrum with a linewidth of more than 100 meV reveals the distribution of sizes and/or chemical composition between the dots in the ensemble. An example of the PL emission of a single CdMnTe QD is displayed in the inset to Fig. 1. As can be seen, the linewidth of this line is of about 4 meV, which is more than an order of magnitude broader than the ~0.1 meV linewidth observed typically for excitons confined to non-magnetic CdTe QDs [32]. This strong broadening of spectral lines is a signature of the formation of magnetic polarons in QDs [30] and is associated with magnetic moment fluctuations within a single QD. The width of the line is related to the binding energy of the EMP [30]. In order to estimate the polaron binding energy we use a simple thermodynamic model, where the emission linewidth, w, should vary as w= $(2\varepsilon_p k_B T)^{1/2}$ [30], where $\varepsilon_p$ is the polaron binding energy, T is the temperature and $k_B$ is the Boltzman constant. Using this model for typical emission lines of width 5meV at T=5K, the polaron binding energy is equal to ~ 30meV, which is significantly larger than binding energies of EMPs observed typically for bulk CdMnTe [19] and CdMnTe QWs [22].

Supporting evidence that the dramatic broadening of single CdMnTe QD emissions is determined by magnetic moment fluctuations is obtained through measurements of the dependence of the linewidth on applied magnetic field. [28]. Since external magnetic fields should suppress the magnetization fluctuations in single CdMnTe QDs, a narrowing of the emission line should also be observed. In Fig. 2 we show the experimentally measured (solid points) full width at half maximum of a single magnetic dot emission line as a function of magnetic field. In the inset, the spectra of a single CdMnTe at B=0T and B=4T (in σ+ polarization) are displayed. Apart from the clear splitting in magnetic field, also substantial narrowing of the emission line is observed. The measured narrowing agrees reasonably with the results of calculation (lines) by Brazis and Kossut [30] performed for two different QD sizes but with constant Mn content of 5%, which assumes the suppression of the magnetic moment fluctuations. It is important to note that by applying magnetic field we also effectively reduce the binding energy of the EMP. However, this reduction in CdMnTe QDs is significantly smaller than for EMP in bulk DMS or even QWs [22]. In such cases, even fields of 0.5 T result in complete alignment of the Mn ions within the EMP [19]. From the absence of complete alignment at fields even up to 4 T for these magnetic QDs, we conclude that this data demonstrates unambiguously the formation of strongly localized zero-dimensional EMPs in this QD sample.

Another important consequence of the spin-spin s,p-d exchange interaction between Mn moments and localized carriers in QDs is the ability to polarize the magnetic ions by optical excitation [11]. As an example, in Fig. 3 (a) we display PL spectra obtained for CdMnTe QDs at T=8K and B=0T. The broad and featureless shaded region represents the PL emission measured for non-resonant excitation just above the ZnTe barrier. In this case, since the photo-excited excitons are randomly captured into the CdMnTe QDs from the ZnTe barrier, the broad PL emission band can be associated with the ground state energy distribution of the QD ensemble. As during this trapping process the carriers go through many scattering events, the excitons randomize their spins before the EMPs are formed. Therefore, for non-resonant excitation, the polarons are randomly aligned and neither net magnetization nor net PL polarization is observed [25].

In order to maintain the spin polarization of the excitons within the formation time of the EMPs we resonantly photo-excite the excitons directly into QD ground states through LO phonon-assisted absorption [33]. In this process, the excitation energy corresponds to the virtual state, which is separated from the actual QD ground state by a multiple of LO phonon energies. In this way the excitons are instantaneously localized by strong QD confinement. Since this strong localization suppresses exciton spin scattering [16,34], one may expect to achieve a very effective transfer of the exciton spin to the Mn ions. This should allow efficient formation of spin-polarized EMPs in these magnetic dots. The resulting PL emission obtained with the resonant excitation is represented by symbols in Fig. 3 (a). As one can see, instead of the broad PL band (see the shaded area), the resonantly excited PL consists of three well resolved LO phonon replicas. Extensive measurements on both large QD ensembles and single non-magnetic

CdTe dots have shown that these three LO phonon replicas are attributed to the QDs that are resonantly excited through the LO phonon assisted absorption [33].

In Fig. 3 (a) we demonstrate how the resonant excitation enables one to control the alignment of the EMPs in the CdMnTe QDs by plotting both $\sigma^+$ (squares) and $\sigma^-$ (circles) – polarized PL spectra obtained at B=0T with $\sigma^+$-polarized excitation. We clearly observe strong predominant $\sigma^+$ polarized PL emission. Moreover, by measuring this polarized PL emission for different excitation energies, we probe the magnetization properties of QDs with different ground state energies, and thus, presumably, with different average sizes. In Fig. 3 (a) we show the polarized PL spectra obtained for two excitation energies of 2.175eV (solid points) and 2.119eV (open points). As one can see, while both cases show significant $\sigma^+$ polarized emission for $\sigma^+$-polarized excitation, the resultant polarization is weaker for the lower excitation energy. Exactly similar results are observed for $\sigma^-$-polarized resonant excitation (not shown). In this case the PL emission is strongly $\sigma^-$ - polarized [11]. These observations re-iterate that using polarized resonant excitation we can optically control the spin-alignment of the EMPs and thereby control the magnetization of the magnetic impurities within the CdMnTe QDs. It is important to emphasize, that the observation of this optical alignment requires that the EMP formation time must be comparable or shorter than the exciton spin relaxation time in the QDs. It may also suggest that, in contrast to bulk DMSs and DMS-based semiconductor QWs, a strong electronic confinement of QDs results in substantial increase of the spin relaxation time of the EMPs.

The polarized PL spectra obtained for two different excitation energies of 2.175 eV and 2.119 eV presented in Fig. 3 (a) show a strong dependence of the total PL intensity of the QDs emission on the excitation energy. For the lower excitation energy (E=2.119 eV, open symbols) the emission intensity is almost an order of magnitude weaker than that excited at E=2.175 eV (solid symbols). Nevertheless, as can be seen, in both cases the QD emission still features strong polarization identical to that of the excitation. To compare directly the degree of polarization for these two excitation energies, in Fig. 3 (b) we calculate the polarization spectra directly from the data in Fig. 2a using the formula $P=(I^+-I^-)/(I^++I^-)$. Here $I^+(I^-)$ are the integrated intensities of $\sigma^+$ ($\sigma^-$) - polarized emissions. We find much higher (more than 50%) polarization when QDs are excited at E=2.175 eV.

In Fig. 4 we show the results of similar analysis for a variety of excitation energies. We plot the polarizations determined by integration of the whole emission band measured at B=0T and at T=8K for $\sigma^+$ - polarized excitations at energies of 2.175eV, 2.156eV, 2.137eV, and 2.119eV. The squares represent the experimental data while the line is a linear fit. The excitation power of the laser was identical for all excitation energies and equal to 2mW. One can see that the polarization of EMPs increases markedly (by 60%) at higher excitation energies. This observation suggests that the optically induced magnetization of the EMPs in CdMnTe depends on the QD energy, and therefore on the average QD size. In other words, the spin polarization of EMPs is stronger for QDs with higher emission energies, thus presumably those with smaller size.

It is important to note that the data shown in Figs. 3 and 4 result from two separate effects. First, the overall intensity of the LO-phonon replicas increases for QDs at higher energies. This requires an increase of the efficiency with which excitons are excited through LO-phonon assisted absorption into the QD ground states. This effect, ascribed by Nguyen *et al.* [33] to the increased exciton-LO phonon coupling in smaller dots, has been discussed in detail for nonmagnetic CdTe dots. Since the average sizes of CdMnTe QDs studied in this work are similar to those in [33] we interpret the variation of the LO phonon replica intensities seen in Fig. 3 (a) as due to enhancement of the exciton-LO phonon coupling for smaller QDs.

On the other hand, we argue that the increase of the PL polarization at higher emission energies does not result from a higher degree of spin-polarization for the resonantly excited excitons. Previously, we have shown [34] that in non-magnetic CdTe QDs as long as excitons are excited through LO-phonon assisted absorption, the degree of polarization of the excitons is the same. In other words, while LO-phonon assisted absorption processes occur more frequently for larger exciton-LO phonon coupling, as long as the resonant excitation occurs, the spin-polarization of the photo-created excitons is equally good. This means that the larger polarization of the PL emission observed for dots at higher energies must be due to a more efficient formation of spin-polarized EMPs. Therefore, we conclude that the increase of the polarization for higher energies of QD emission seen in Fig. 4 is a result of the dependence of the EMP properties on the QD size. In fact, as it has been shown theoretically [26] the binding energy of EMP is larger for smaller CdMnTe dots. One may also expect that the spin-polarization of Mn ions depends on the stability of the EMPs. Thus, we speculate the increased spin-polarization of EMPs at higher energy might reflect the increased stability of the optically induced magnetization of EMPs for smaller QDs.

As shown in Fig. 1, the very broad emission line of a single CdMnTe QD reflects the very large binding energy of the EMPS in these structures. One might thus expect that the photo-induced magnetization of EMPs observed for these QDs should persist up to very high temperatures. In Fig. 5 we show the polarization of the zero-field PL as a function of temperature. The data have been obtained with $\sigma^+$ - polarized excitation at 2.175eV with temperature ranging from 8K and 120K. The experimental data measured only for the first LO phonon replica is represented by the squares with the linear fit shown by a solid line. The extrapolation of the fit shows that the spin-polarization of the EMP confined to CdMnTe QD persists up to 160K. We believe that the enhanced thermal stability of the photo-induced magnetization reflects the large (confinement – enhanced) binding energy of the zero-dimensional EMPs in the magnetic QDs (see Fig. 1).

Previous experiments performed on Mn-doped bulk semiconductors [9] have revealed a strong increase of the optically induced magnetization with increasing excitation power of the circularly polarized laser. This effect has been attributed to much more efficient spin transfer between larger number of photo-created carriers and Mn ions. In fact, since in bulk or QW DMS there is no immediate localization of the excitons, as estimated by Awschalom *et al.*, only 1 per million of the available Mn impurities could be ferromagnetically aligned through optical excitation [8]. This relatively low rate is clearly due to spin relaxation processes that occur during self-localization of carriers that eventually leads to the formation of magnetic polarons. On the other hand, in QDs the spin-polarized carriers are photo-excited directly into the ground states of QDs. Therefore any diffusion of carriers can be neglected and only the Mn impurities confined to QDs can be ferromagnetically aligned. The results of excitation power dependence of the polarization of the QD emission are shown in Fig. 6. All the spectra were taken for $\sigma^+$ - polarized excitation at the laser energy of 2.156eV and the excitation power was equal to (a) 0.4mW and (b) 4mW. The PL spectra are normalized to the maximum of the $\sigma^+$ - polarized first LO phonon line. In addition, we also show the polarization curves calculated for these two datasets. As can be seen, a change in the excitation power by factor of 10 does not influence the measured polarization. The insensitivity of the optically induced magnetization to the number of photo-excited carriers suggests that the polarization of the EMPs is already complete at the lowest excitation powers. We attribute this result to the immediate strong localization of the resonantly excited excitons in the CdMnTe QDs.

Apart from being influenced by the QD size or the sample temperature, the optically induced magnetization depends also on the manganese concentration in the QD sample. Figure 7 displays the resonantly excited PL spectrum of the as grown QD samples with (a) low and (b) high Mn content obtained at B=0T. Estimated average concentrations of magnetic impurities are 2% and 4.5%, respectively. We also show the polarized spectra measured of the annealed CdMnTe QD sample (Fig. 7 (c)). It is important to note that although the PL spectra were obtained with different excitation energies (as the energy distribution of QDs varies between the samples), the laser energy in each case is tuned to approximately 1LO phonon above the maximum of the non-resonant PL emission. This allows more direct comparison of the polarization of the QD emission, as any possible effects of change in the exciton-LO phonon coupling are minimized. The polarization spectra calculated from these data are also displayed in Fig. 7. As one can see for the as-grown samples, the overall polarization is significantly higher for QDs with lower Mn content (see Fig. 7 (a)). We attribute this effect to weaker contribution from antiferromagnetic Mn-Mn nearest neighbor interaction [6] in the QDs with lower average Mn concentration. Such an interaction prevents the complete polarization of the EMP and also allows the EMP, once created, to depolarize more effectively. Interestingly, the optically induced magnetization of the QD sample with high Mn concentration features a strong increase after rapid thermal annealing (see Fig. 7 (c)). It has been shown recently that annealing increases the average size of the QDs but also may induce interdiffusion of Mn out of QDs [25]. Since, presumably, both these effects would have similar qualitative impact on the Mn-Mn interaction, we cannot at present distinguish which of the two plays dominant role. We believe that experiments on single CdMnTe QDs should allow separation of these two contributions. Nevertheless, the results displayed in Fig. 7 demonstrate a useful way of tuning the spin polarization of the magnetic impurities in self-assembled QDs.

In summary, we demonstrate that for resonant excitation of spin-polarized excitons into CdMnTe magnetic quantum dots, significant magnetization of the Mn-spins can be obtained through the formation of polarized exciton magnetic polarons. We find that this Mn polarization persists to temperatures as high as 160 K, reflecting the strong stability of the exciton magnetic polarons in these QD structures. The detailed ensemble studies of the dependence of the optically induced magnetization on both QD size and Mn concentration provide further important insights into the nature of this effect. We believe that ongoing single dot experiments accompanied with theoretical efforts should make possible a comprehensive description of both dynamics and the equilibrium characteristics of ferromagnetically aligned magnetic quantum dots.

The work was supported by the National Science Foundation through 9975655 and 0071797 (USA) the State Committee for Scientific Research under Grant PBZ-KBN-044/P03/2001 (Poland).

**Figure Captions:**

Figure 1. Low temperature PL spectrum of the ensemble of CdMnTe QDs excited non-resonantly, above the ZnTe barrier. In the inset, typical emission lines of three single magnetically – doped quantum dots at T=6K.

Figure 2. Magnetic field dependence of the full width at half maximum of a single CdMnTe QD emission at T=6K. Symbols represent the experimental data while the lines are the result of calculation for two quantum dot sizes of 2nm and 3nm. In the inset the spectra measured at B=0T and at B=4T are shown.

Figure 3. (a) Resonantly excited PL spectra of CdMnTe QDs measured at B=0T and T=8K for two excitation energies of 2.175 eV (solid symbols) and 2.119 eV (open symbols). The excitation was $\sigma^+$ - polarized while both $\sigma^+$ (squares) and $\sigma^-$ (circles) - polarized components of the emission were analyzed. The shaded region represents the PL emission for non-resonant excitation just above the ZnTe barrier. (b) Polarization calculated for the spectra shown in Fig. 2a. Solid (open) points represent the data measured at the excitation energy of 2.175 eV (2.119 eV).

Figure 4: Excitation energy dependence of the PL polarization (squares) of CdMnTe QDs measured at B=0T and T=8K. The excitation power is kept constant at 2 mW. The line is the linear fit to the data.

Figure 5. Temperature dependence of the polarization of the PL emission associated with the first LO phonon absorption measured of CdMnTe QDs at B=0T. The excitation was $\sigma^+$ - polarized. The squares represent the experimental data and the line is a linear fit.

Figure 6. Resonantly excited PL spectra of CdMnTe QDs measured for $\sigma^+$ - polarized excitation at energy 2.156 eV together with respective polarization plots. The data obtained with the excitation power of (a) 4 mW and (b) 0.4 mW are shown.

Figure 7. Resonantly excited PL spectra of three different CdMnTe QDs measured at T=6K for $\sigma^+$ - polarized excitation together with respective polarization plots. (a) as-grown CdMnTe QDs with low (~2%) average Mn concentration (b) as-grown CdMnTe QDs with high (~4.5%) average Mn concentration (c) sample (b) but annealed for 15 seconds at T=420C.

T. Gurung, et al., Figure 1

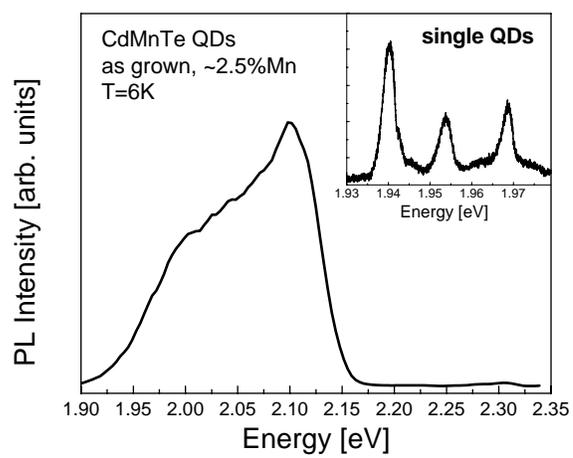

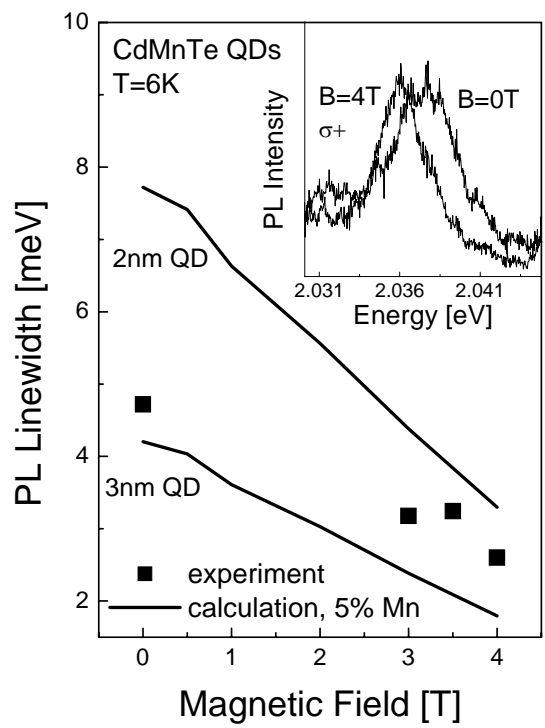

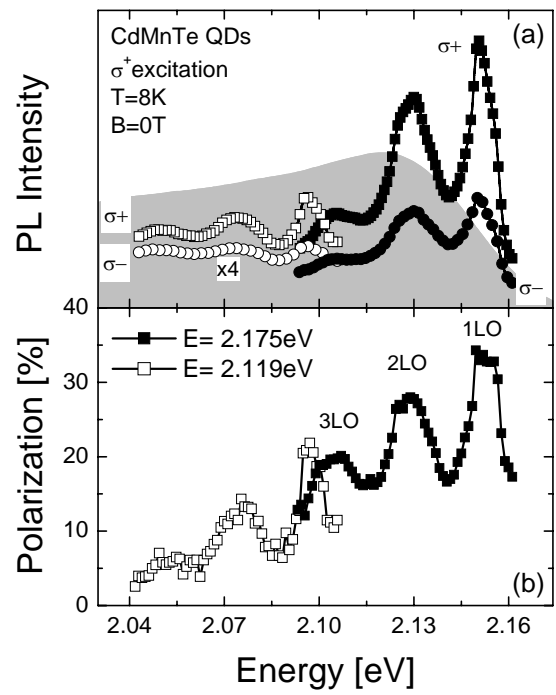

T. Gurung, et al., Figure 4

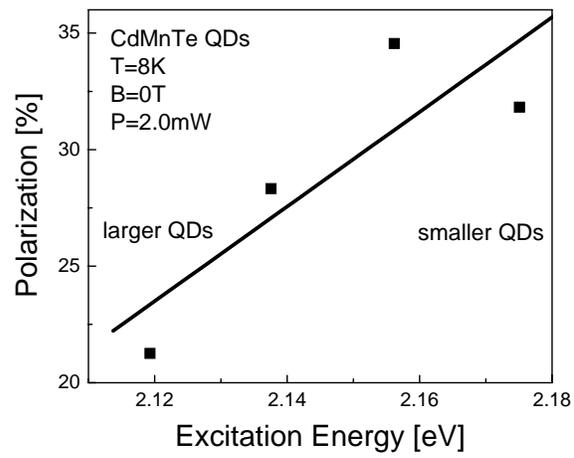

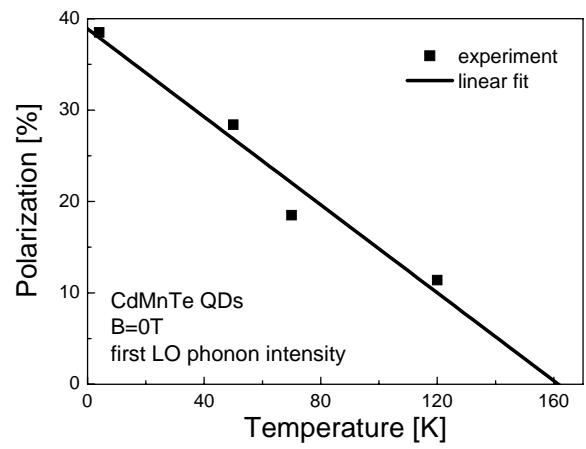

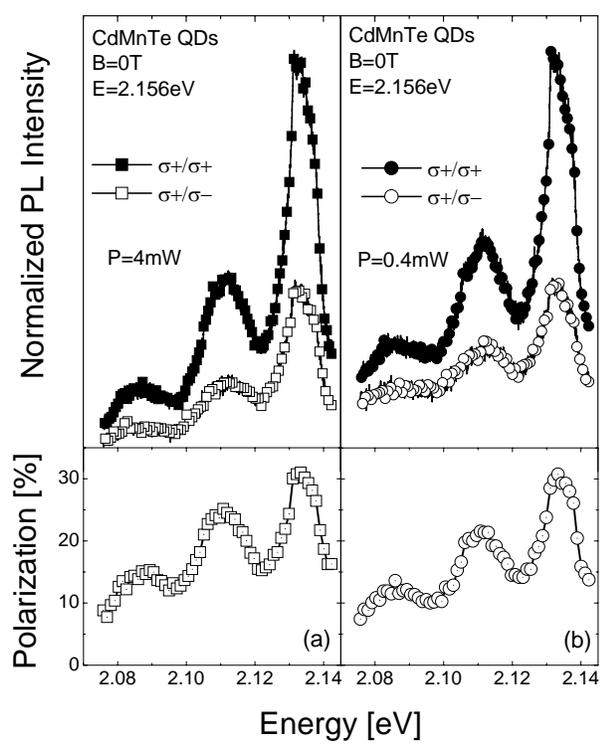

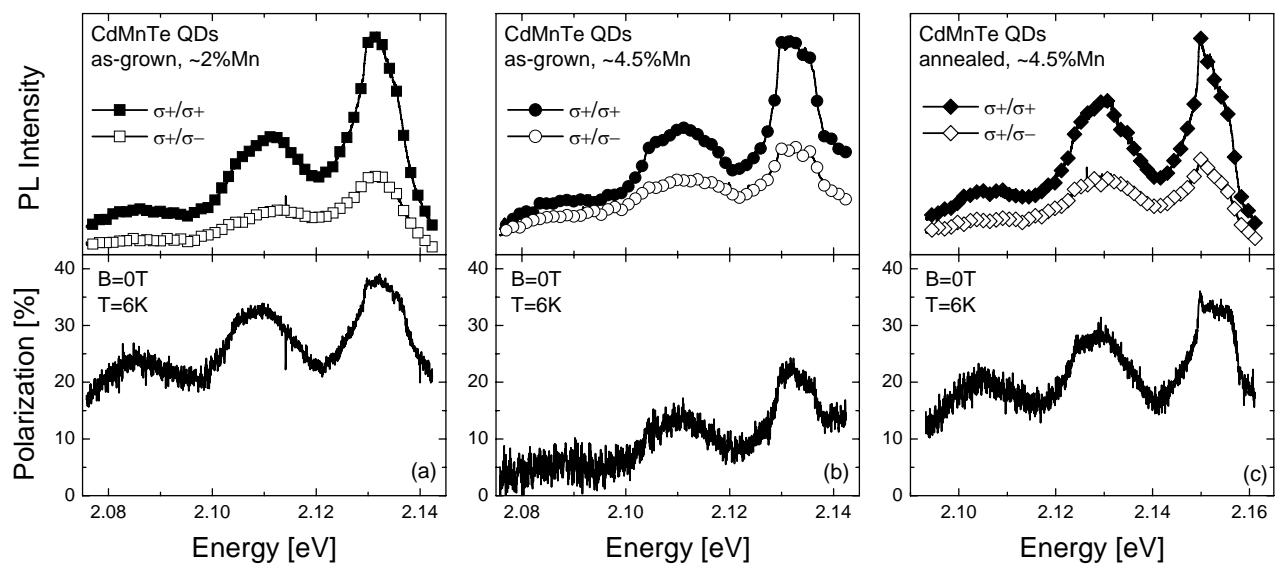